\documentclass{PoS}
\usepackage[normal]{subfigure}

\title{Ionization and maximum energy of nuclei in shock acceleration theory}

\ShortTitle{Ionization and maximum energy in shock acceleration theory}

\author{Giovanni Morlino\\
        INAF - L.go E. Fermi 5,  Firenze, Italy\\
        E-mail: \email{morlino@arcetri.astro.it}}

\abstract{We study the acceleration of heavy nuclei at SNR shocks when the
process of ionization is taken into account. Heavy atoms ($Z_N >$ few) in the
interstellar medium which start the diffusive shock acceleration (DSA) are never
fully ionized at the moment of injection. The ionization occurs during the
acceleration process, when atoms already move relativistically. For typical
environment around SNRs the photo-ionization due to the background galactic
radiation dominates over Coulomb collisions. 
The main consequence of ionization is the reduction of the maximum energy which
ions can achieve with respect to the standard result of the DSA. In fact the 
photo-ionization has a timescale comparable to the beginning of the 
Sedov-Taylor phase, hence the maximum energy is no more proportional to the
nuclear charge, as predicted by standard DSA, but rather to the effective ions'
charge during the acceleration process, which is smaller than the total nuclear
charge $Z_N$. This result can have a direct consequence in the prediction of
the {\it knee} structure of the cosmic ray spectrum. Moreover the acceleration
of ultra-heavy elements beyond the Iron's maximum energy is very hard to achieve
making unlikely their possible contribution to the cosmic ray spectrum in the
transition region between Galactic and extragalactic component.}

\FullConference{25th Texas Symposium on Relativistic Astrophysics - TEXAS 2010\\
		December 06-10, 2010\\
		Heidelberg, Germany}

\begin{document}

\section{Introduction}

The bulk of Galactic cosmic rays (CRs) is largely thought to be accelerated at
the shock waves associated with supernova remnants (SNRs) through the mechanism
of diffusive shock acceleration (DSA). A key aspect of DSA is the maximum
energy achieved by different nuclei which is intimately connected with the
interpretation of the {\it knee} structure and of the transition region from
Galactic to extragalactic CR component. The {\it knee} is commonly interpreted
as due to the superposition of the spectra of all chemicals with different
cutoff energies. Using the flux of different components measured at low energies
it has been shown that the {\it knee} is well reproduced if one assumes that the
maximum energy of each specie $E_{\max,N}$ is proportional to the nuclear charge
$Z_N$ \cite{horandel03}.
Nevertheless the superposition of subsequent cutoff seems to be confirmed by the
measurements of the spectrum of single components in the {\it knee} region: data
presented by the KASCADE experiments show that the maximum energy of He is $\sim
2$ times larger than that of the protons \cite{antoni05}.
The relation $E_{\max,N} \propto Z_N$ is clearly predicted by DSA if one assume
that the diffusion coefficient is rigidity-dependent and that nuclei are
completely ionized during the whole acceleration process. Indeed the second
assumption does not hold in general. When ions are injected into the
acceleration process they are unlikely to be fully ionized, especially if of
high nuclear charge. In fact the temperature of the circumstellar medium where
the forward shock propagates varies from $10^4$ up to $10^6$ K. If $T \sim
10^4$ K even hydrogen is not fully ionized, as demonstrated by the presence of
Balmer lines associated with shocks in some young SNRs \cite{Heng09}. For $T
\sim 10^6$ only atoms up to  $Z_N= 5$ can be completely ionized.
The ordinary assumption made in the literature is that atoms are completely
stripped soon after the beginning of the acceleration process. In spite of this
assumption in \cite{mor09} we showed that, for a typical SNR shock, the
ionization time is comparable with the acceleration time, hence electrons are
stripped long time after the injection, when ions already move
relativistically. 
This fact has two important consequences: 1) the maximum energy of ions can be
reduced with respect to the standard prediction of DSA and 2) the ejected
electrons can easily start the acceleration process providing a source for the
synchrotron radiation. The latter point has been already analyzed in
\cite{mor09} and \cite{mor11}. Here we want to describe how the ionization can
affect the maximum energy achieved by heavy ions during the acceleration at SNR
shocks.

\section{Ionization vs. acceleration time} \label{sec:time}
In order to get the maximum energy achieved by different chemical specie, we
need to compare the acceleration time with the ionization time.
Let us consider atoms of a single specie $N$ with nuclear charge $Z_N$ and mass
$m_N= A m_p$, which start the DSA with initial charge $Z<Z_N$ and momentum
$p_{\rm inj}$. We adopt the acceleration time as computed in the framework of
linear DSA for plane shock geometry:
\begin{equation}
 \tau_{\rm acc}(p_{\rm inj}, p)= \int_{p_{\rm inj}}^{p}\frac{3}{u_1-u_2} \left(
  \frac{D_1(p)}{u_1} +
  \frac{D_2(p)}{u_2} \right) \frac{dp}{p} 
 = 0.85\, \frac{\beta\, (p-p_{\rm inj})}{m_N c} \, B_{\,\mu G}^{-1} \, u_8^{-2} 
    \left( \frac{A}{Z} \right) \,{\rm yr} \,.
 \label{eq:t_acc1}
\end{equation}
Here $u$ is the plasma speed in the shock rest frame, and the subscript 1
(2) refers to the upstream (downstream) quantities (note that $u_{\rm shock}
\equiv u_1$). The downstream plasma speed is related to the upstream one
through the compression factor $r$, i.e. $u_2 = u_1/r$. We limit our
considerations to strong shocks, which have $r=4$. The turbulent magnetic field
responsible for the particle diffusion is assumed to be compressed downstream
according to $B_2= r\,B_1$. The last equality in Eq.~(\ref{eq:t_acc1}) holds
because we assume Bohm diffusion coefficient, that is $D_B=r_L \beta c/3$, where
$\beta c$ is the particle speed and $r_L= pc/Z eB$ is the Larmor radius. 
Here the magnetic field is expressed in $\mu$G and the shock speed is $u_1= u_8
10^8 {\rm cm/s}$. 

Eq.~(\ref{eq:t_acc1}) has to be compared with the ionization time.
Ionization can occur either via Coulomb collisions with thermal particles or via
photo-ionization with background photons. In \cite{mor11} we showed that
Coulomb collisions are negligible when the density of the circumstellar medium
is $n_1 \lesssim 30\,{\rm cm^{-3}}$. This condition is satisfied for almost all
SNR environments, with the exception of the very initial phase of the explosion,
when the shock propagates in the dense progenitor's wind, or when the remnant
collides with a molecular cloud. Here such scenarios will be neglected.

The full ionization time results from the convolution of the photo-ionization
cross section with the photon energy spectrum, $dn_{\rm ph}/d\epsilon$, due to
the total interstellar radiation field (ISRF), i.e.:
\begin{equation} \label{eq:Ph_time}
 \tau_{\rm ph}^{-1}(\gamma)= \int d\epsilon\, 
 \frac{dn_{\rm ph}(\epsilon)}{d\epsilon} \,c\,
     \sigma_{\rm ph}(\gamma \epsilon)  \,,
\end{equation}
where $\epsilon'= \gamma \epsilon$ is the photon energy as seen in the ion rest
frame, which moves with Lorentz factor $\gamma$. The photo-ionization cross
section can be estimated using the simplest approximation for the $K$-shell
cross section of hydrogen-like atoms with effective nuclear charge $Z$, i.e. 
$
 \sigma_{\rm ph}(\epsilon') = 64\,\alpha^{-3}\sigma_T Z^{-2}
  \left( I_{N,Z}/\epsilon'\right)^{7/2} 
$
\cite{heitler}, where $\sigma_T$ is the Thompson cross section, $\alpha$ is the
fine structure constant and $I_{N,Z}$ is the ionization energy threshold for the
ground state of the chemical specie $N$ with $Z_N-Z$ electrons. The numerical
values of $I_{N,Z}$ can be found in the literature (see e.g. \cite{allen73}).  

\section{Maximum energy of Ions}  \label{sec:E_max}
We stress that a consistent treatment of the ionization effects would require
the use of time-dependent calculation. However, in order to get an approximate
result for the maximum energy we can use the quasi-stationary version of the
linear acceleration theory.

First of all we assume that the maximum energy is achieved at the beginning of
the Sedov-Taylor phase ($t_{ST}$) \cite{BAC-pmax07}. In fact, for later times,
$t>t_{ST}$, the shock speed decreases faster than the diffusion velocity, hence
particles at the maximum energy can escape from the accelerator and the maximum
energy cannot increase further. If the total ionization time is comparable, or
even larger than $t_{ST}$, we do expect that $E_{\max} < E_{\max}^0$, where we
call $E_{\max}^0$ the maximum energy achieved by ions which are completely
ionized since the beginning of acceleration.
We also assume that initial level of atoms' ionization is only determined by
the circumstellar medium temperature. We neglect complications arising from the
dust sputtering process which could play an important role in the injection of
ions in the DSA mechanism \cite{ellison97}.

Now let us consider a single ion injected with momentum $p_{\rm inj}$ and total
charge $Z_1$. The ion undergoes acceleration at a constant rate $\propto Z_1$
during a time equal to the ionization time needed to lose one electron,
$\tau_{\rm ph,1}$, when it achieves the momentum $p_1$. $\tau_{\rm ph,1}$ and
$p_1$ can be determined simultaneously equating the acceleration time with the
ionization time, i.e. $\tau_{\rm ph,1}(p_1/m_Nc) = \tau_{\rm acc}(p_0,p_1)$.
Using Eq.~(\ref{eq:t_acc1}) the last condition gives:
\begin{equation}
 \frac{p_1}{m_p c}= \frac{p_{\rm inj}}{m_p c} + \frac{Z_1 B_1 u_1^2}{1.7 \, Z_N}
                    \, \tau_{\rm ph,1}\left(p_1/m_N c\right) \,,
 \label{eq:p2_sol}
\end{equation}
where the subscript $1$ label the quantities during the time $\tau_{\rm ph,1}$.
Because the photo-ionization occurs when ions already move relativistically
\cite{mor09}, we set $\beta=1$. In Eq.~(\ref{eq:p2_sol}) $B$ is expressed in
$\mu G$ and $u_{sh}$ in units of $10^8$ cm/s, while $\tau_{\rm ph,1}$ is
expressed in yr. 
Once the background photon distribution is known Eq.~(\ref{eq:p2_sol}) can be
solved numerically. After the first ionization the acceleration proceeds at a
rate proportional to $Z_2 \equiv Z_1+1$ during a time needed to lose the second
electron, $\tau_{\rm ph,2}$. Applying Eq.~(\ref{eq:p2_sol}) repeatedly for all
subsequent ionization steps, we get the momentum when the ionization is
complete. If the total ionization time, $\tau_{\rm ion}^{\rm tot}$, is
smaller than the Sedov time, in order to get the maximum momentum we need to add
the further acceleration during the time $(t_{ST} - \tau_{\rm ion}^{\rm tot})$,
with a rate $\propto Z_N$. The final expression for the maximum momentum can be
written as follows:
\begin{equation}
 \frac{p_{\max}}{m_pc}=  \frac{p_{\rm inj}}{m_p c} + \sum_{k=1}^{Z_N-Z_0} 
      \frac{Z_k B_k u_k^2 \tau_{{\rm ph},k}}{1.7\, Z_N}
    + \theta\left(t_{ST}-\tau_{\rm ion}^{\rm tot}\right)
      \sum_{i=1}^{M} \frac{B_i u_i^2}{1.7\, M}\,,
 \label{eq:p2_sol2}
\end{equation}
The last term has been written as a sum over $M$ time-steps in order to handle
the case where magnetic field and shock speed change with time.

Let us analyze first the stationary case, with $u_{sh}$ and $B$ constant
during all the free expansion phase. We consider two different situations which
can represent a type I/a and a core-collapse supernova (CC SNa). Type I/a SNe
explode in the regular interstellar medium whose typical density and
temperature are $n_1= 1$ cm$^{-3}$ and $T_0= 10^4$ K.  Conversely SNRs generated
by CC SNe expand into a diluted and hot bubble generated either by the
progenitor's wind or by the explosion of other close SNe (so called
\textit{super-bubble})\cite{higdon}. In both cases typical values for the
density and temperature inside the bubble are $n_1= 10^{-2}$ cm$^{-3}$ and $T_0=
10^6$ K. 
For both type I/a and CC SNe we assume the same value for the explosion energy
$E_{SN}= 10^{51}$ erg, and mass ejecta $M_{ej}= 1.4 M_{\odot}$. The resulting
Sedov times are $t_{ST}= 470$ yr and $t_{ST}= 2185$ yr, respectively. The
average shock speed during the free expansion phase is $u_{sh}\simeq 6000$ km/s
in both cases.

The last parameter we need to estimate is the magnetic field strength which can
be inferred assuming that both kind of SNe are able to accelerate protons up to
the {\it knee} energy, which is $E_{\rm knee}= 3\cdot 10^{15}$ eV. This
condition gives $B_1= 160 \mu G$ and $35 \mu G$ for type I/a and CC cases,
respectively. It is worth noting that the chosen values of $B_1$ are consistent
with those predicted by the CR-induced magnetic filed amplification.

\begin{figure}
 \begin{center}
 \subfigure[Type I/a: stationary]
   {\includegraphics[width=7cm]{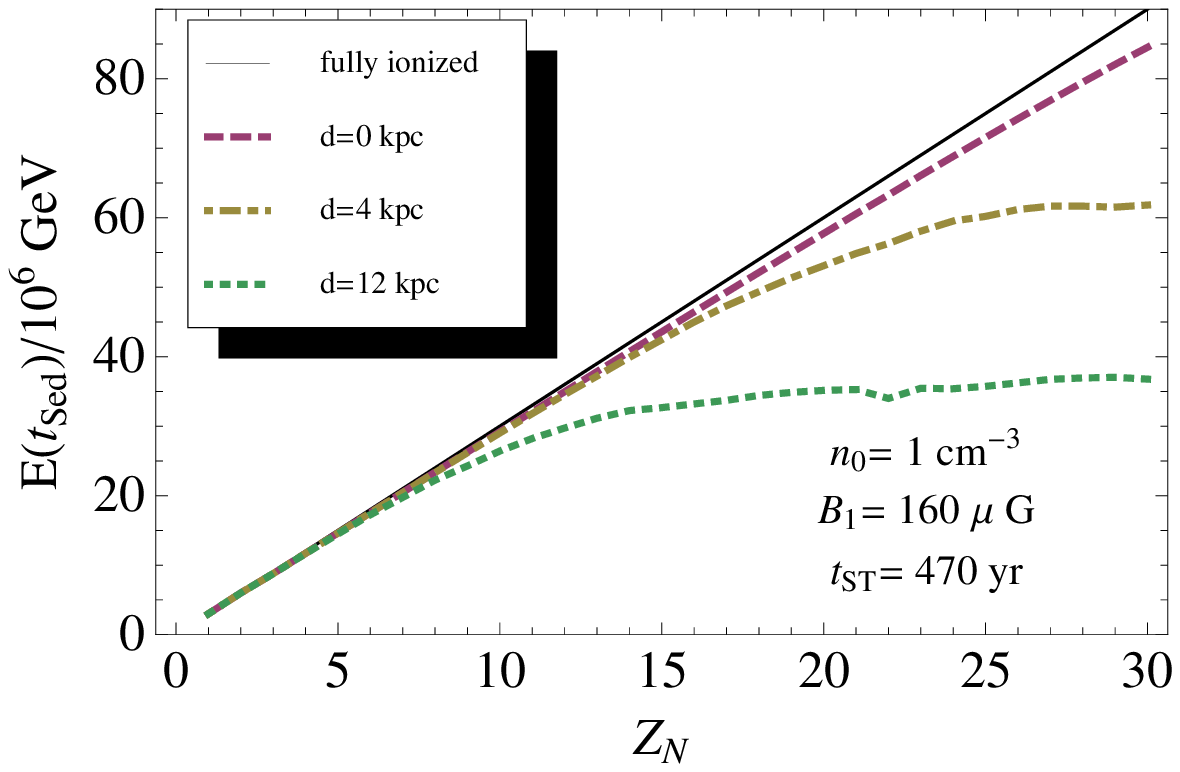}}
 \hspace{5mm}
 \subfigure[Core-collapse: stationary]
   {\includegraphics[width=7cm]{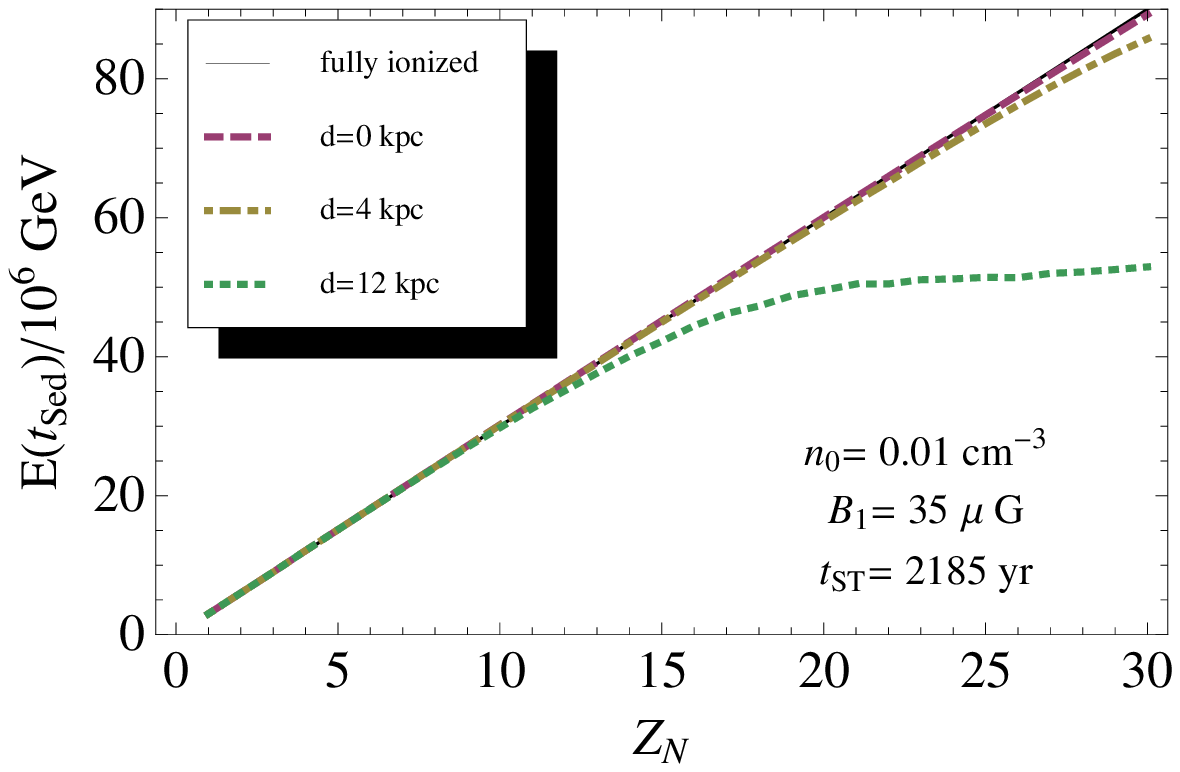}}
\end{center}
\begin{center}
 \subfigure[Type I/a: evolution]
   {\includegraphics[width=7cm]{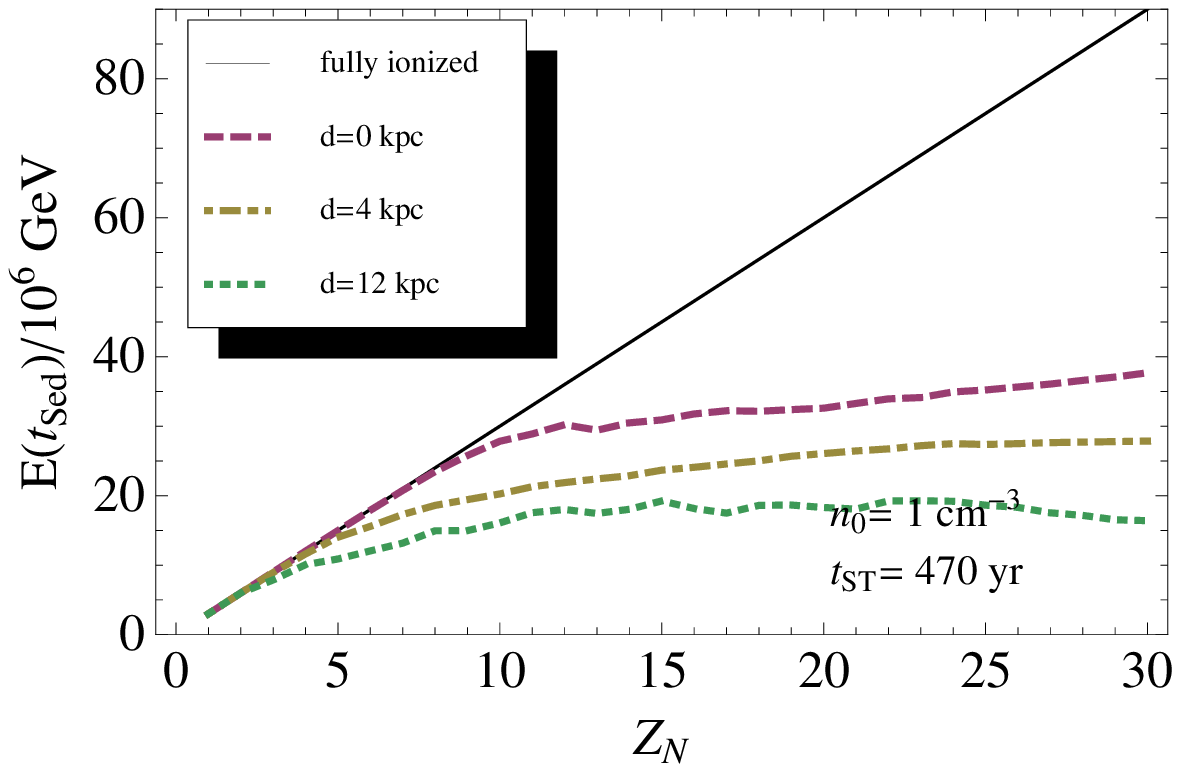}}
 \hspace{5mm}
 \subfigure[Core-collapse: evolution]
   {\includegraphics[width=7cm]{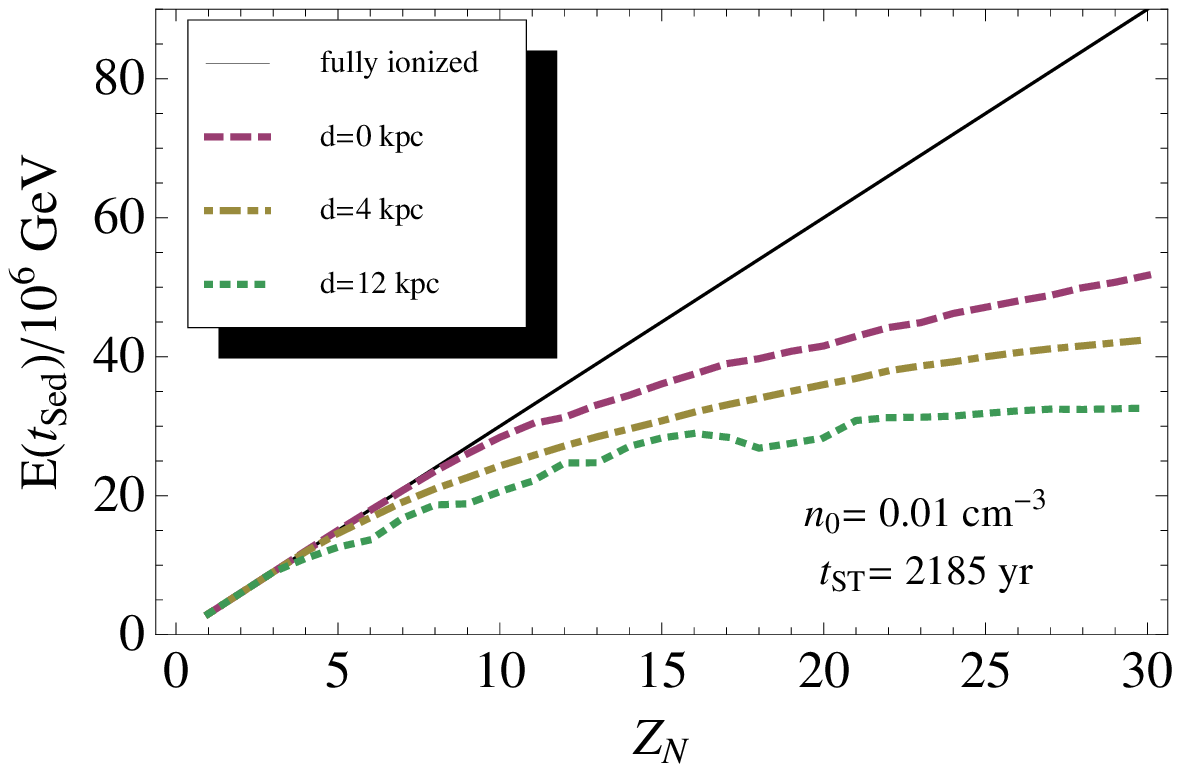}}
\end{center}
 \caption{Maximum energy achieved by chemical specie with nuclear charge $Z_N$
(ranging from H to Zn) at the beginning of the Sedov phase. Panels (a) and (b)
show the cases of type I/a and core-collapse SNR for a constant shock speed and
magnetic field strength, while for panels (c) and (d) $u_{\rm sh}$ and $B$
evolve with time, as explained in the text. The thin solid line shows the
maximum energy achieved by atoms which are fully ionized since the beginning of
the acceleration, while the other curves are computed including the
photo-ionization due to ISRF for SNRs located at three different distance from
the Galactic Center, namely 0, 4 an 12 kpc.}
\label{fig:Emax}
\end{figure}

The effectiveness of ionization dependents on the ISRF which is a decreasing
function of the distance from the Galactic Center. We adopt the ISRF as computed
in \cite{porter05}.
In Fig.~(\ref{fig:Emax}) we plot the maximum energy achieved for $t=t_{ST}$ by
different chemical species, from H up to Zn ($Z_{\rm Zn}=30$). The panels (a)
and (b) show the case of type I/a and CC SNRs, respectively. Each panel contains
four lines: thin solid lines are the maximum energy achieved by ions which start
the acceleration completely stripped, $E_{\max}^0$, while the remaining lines
show $E_{\max}$ computed according to Eq.~(\ref{eq:p2_sol2}) for three different
locations of the SNR in the Galactic plane: in the Galaxy Center and at 4 and 12
kpc away from it. Looking at the panel (a) we see that the maximum energy
achieved by different nuclei in type I/a SNRs does not increase linearly with
$Z_N$, instead it reaches a plateau for $Z_N \gtrsim 25$ at a distance $d=4$ kpc
and for $Z_N \gtrsim 15$ at $d= 12$ kpc. Only for SNRs located in the Galactic
bulge the proportionality relation $E_{\rm max}\propto Z_N$ holds, at least for
elements up to $Z_N= 30$. The effect of ionization is much less relevant for
core collapse SNRs: only those remnants located at a distance of 12 kpc show a
noticeable reduction of $E_{\max}$.

Previous conclusions are based on the assumption that both the magnetic field
and the shock velocity remain constant. This is indeed a poor approximation. In
fact during the free expansion phase the shock speed can vary by a factor of
few. Now for DSA the acceleration rate is proportional to $u_{sh}^2 \delta
B(u_{sh})$, where $\delta B(u_{sh})$ is the turbulent magnetic field generated
by the CR-induced instabilities (resonant or non-resonant), which is, in turn,
an increasing function of $u_{sh}$. Hence the acceleration rate can
significantly change during the free expansion phase. In order to evaluate the
effect of the time evolution we can use a simple toy model always in the
framework of steady-state approach: we approximate the continuum evolution
into time steps equal to the ionization times, $\tau_{\rm ph,k}$, assuming that
the stationary approximation is valid during each time step. For the description
of the shock dynamic we follow \cite{truelove99}. Specifically we adopt the
solution for a remnant characterized by a power-law profile of the ejecta with
index $n=7$, expanding into an homogeneous medium. For this specific case the
shock velocity for $t<t_{ST}$ is well described by $u_1(t)/u_{ch}= 0.606\,
(t/t_{ch})^{-3/7}$ (see Table~7 in \cite{truelove99}) where
$u_{ch}=(E_{SN}/M_{ej})^{1/2}$, $t_{ch}= E_{SN}^{-1/2} M_{ej}^{5/6}
\rho_1^{-1/3}$ and $t_{ST}= 0.732 \,t_{ch}$. We adopt the same values of
$E_{SN}$, $M_{ej}$ and $\rho_1$ used in the cases (a) and (b).

For what concern the magnetic field we assume that the amplification mechanism
converts a fraction of the incoming kinetic energy flux into magnetic energy
density downstream of the shock: $\delta B_2^2(t)/(8\pi)= \alpha_B \rho_1
u_1^2(t)$.
The parameter $\alpha_B$ hides all the complex physics of magnetic amplification
and particle acceleration. For the sake of completeness we mention that in the
case of resonant streaming instability $\alpha_B \propto \xi_{cr} v_A/u_1$,
where $\xi_{cr}$ is the efficiency in CRs and $v_A$ is the Alfv\'en velocity,
while in the case of resonant amplification $\alpha_B \propto \xi_{cr} u_1/4c$
\cite{bell04}.
However such relations cannot be applied in a straightforward way since they
require using a non linear theory. Here we prefer to make the simplest
assumption, taking $\alpha_B$ as a constant. As we done for the stationary
case, we determine the value of $\alpha_B$ assuming that for $t=t_{ST}$ the
energy achieved by protons is $3\cdot 10^{15}$ eV. This condition gives
$\alpha_B= 3.25 \cdot 10^{-3}$ and $6.80\cdot 10^{-3}$ for type I/a and CC
cases, respectively. 
It is worth noting that in the case of some young SNRs the value of $\alpha_B$
has been estimated from the measurement of both the shock speed and the magnetic
field strength and the results are only a factor $5-10$ larger then the values
we use here (see, e.g., Table 1 from \cite{caprioli09}).

Now Eq.~(\ref{eq:p2_sol2}) can be used to evaluate the effect of time
evolution.  For each time-step, $\tau_{{\rm ph},k}$, the values of $u_k$ and
$B_k$ are computed according to the equation described above, evaluated at the
beginning of the time-step. Always in Fig.~\ref{fig:Emax} we report the results
for the maximum energy at $t= t_{ST}$ for the case of type I/a and CC SNRs in
panels (c) and (d), respectively. Our results show that the reduction of the
maximum energy is now more pronounced with respect to the stationary case shown
in the panels (a) and (b). Even in the scenario of a core-collapse SNR located
in the Galactic bulge, iron nuclei are accelerate only up to about one half of
the maximum theoretical energy.
The reason why ionization is less effective when the evolution is taken
into account is a consequence of the fact that acceleration occurs mainly during
the first stage of the SNR expansion. In fact both the shock speed and the
magnetic field strength are larger at smaller times. On the other hand the
effective ions' charge is small during the initial phase of the expansion,
hence the acceleration rate is smaller than its maximum possible value.

\section{Discussion and conclusions}  \label{sec:conclusion}
The most relevant consequence of the ionization mechanism concerns the shape of
the {\it knee} in the CR spectrum. As we have already discussed, the relation
$E_{\max,N} \propto Z_N$ is needed in order to fit the data in the {\it knee}
region, which has a slope equal to 3.1. Even a small deviation of the cutoff
energy from the direct proportionality can affect this prediction.
Our results suggest that type I/a SNRs seem unable to accelerate ions up to an
energy $Z_N$ times the proton energy, due to their small Sedov-Taylor age. This
consideration is strengthened by the fact that the CR flux observed at the Earth
is mostly due to the SNRs located in the solar neighborhood, rather than those
located in the Galactic bulge\footnote{In fact the escaping length from the
Galaxy is determined by the thickness of the Galactic halo which is $\sim 3-5$
kpc, hence particles can reach the Earth only if they are produced within this
distance, while the Galactic bulge is at 8 kpc from us.} where the larger photon
field make the ionization faster and allow nuclei to achieve larger energies. 
Ions accelerated at SNRs produced by CC SNe are less affected by the ionization
and can achieve larger energies because the larger Sedov-Taylor age. But even
in this case the maximum energy could be appreciably reduced  when the time
evolution of the remnant is taken into account.
We can conclude that the primary sources of CRs above the {\it knee} energy are
most probably the core-collapse SNRs with $M_{ej}\gg 1 M_{\odot}$. 

A second comment concerns the acceleration of elements beyond the iron group.
Even if the contribution of such ultra-heavy elements to the CR spectrum is
totally negligible at low energies, in the 100 PeV regime it could be
significant. This region is where the transition between Galactic and
extragalactic CRs occurs. Indeed several authors pointed out that a new
component is needed to fit this transition region, beyond the elements up to
iron accelerated in ``standard'' SNRs (see e.g. the discussion in
\cite{caprioli10}). As inferred by \cite{horandel03} stable elements heavier
than iron can significantly contribute to the CR spectrum in the 100 PeV regime
if one assume that their maximum energy scales like $Z_N$. This assumption is
especially appealing also because it could explain the presence of the {\it
second knee} in the CR spectrum \cite{horandel03}. In fact the ratio
$E_{2^{nd}knee} / E_{knee} \simeq 90$ is very close to the nuclear
charge of uranium, which should have $E_{\max,U}= 92 E_{\max,p}= 414$ PeV.
On the other hand  the ionization mechanism discussed here provides a strong
constraint on the role of ultra-heavy elements. If the acceleration of elements
heavier than Fe occurs at SNRs like those considered in this work, it is
easy to show that they cannot achieve energies much larger than the Fe itself,
because the total ionization time increases rapidly with the nuclear charge. 
Even if we consider the acceleration in very massive SNRs, with $M_{ej}\gg 1
M_{\odot}$, the contribution of ultra-heavy elements remains unlikely. 
An interesting alternative worth to investigate is the acceleration during the
very initial stage of the explosion, when the expansion occurs into the dense
progenitor's wind and the Coulomb collisions can strongly reduce the ionization
time.

\end{document}